\documentclass[reprint,amsmath,amssymb,aps,prb,superscriptaddress,longbibliography]{revtex4-2}

\usepackage{array}
\usepackage{bm}
\usepackage{color}
\usepackage{float}
\usepackage{graphicx}
\usepackage{hyperref}
\hypersetup{colorlinks=true,allcolors=blue}
\usepackage{natbib}
\usepackage{newtxtext}
\usepackage{newtxmath}
\usepackage{amssymb}
\usepackage{ulem}
\usepackage[caption=false]{subfig}

\begin{document}

\title{Correlation Matrix Method for Phonon Quasiparticles}

\author{Wenjing Li}
\affiliation{Key Lab of Advanced Optoelectronic Quantum Architecture and Measurement (MOE), Beijing Key Laboratory of Quantum Matter State Control and Ultra-Precision Measurement Technology, and School of Physics, Beijing Institute of Technology, Beijing 100081, China}
\affiliation{International Center for Quantum Materials, Beijing Institute of Technology, Zhuhai, 519000, China}

\author{Yong Lu}
\affiliation{College of Mathematics and Physics, Beijing University of Chemical Technology, Beijing 100029, China.}

\author{Fawei Zheng}
\email[]{Author to whom correspondence should be addressed:  fwzheng@bit.edu.cn}
\affiliation{Key Lab of Advanced Optoelectronic Quantum Architecture and Measurement (MOE), Beijing Key Laboratory of Quantum Matter State Control and Ultra-Precision Measurement Technology, and School of Physics, Beijing Institute of Technology, Beijing 100081, China}
\affiliation{International Center for Quantum Materials, Beijing Institute of Technology, Zhuhai, 519000, China}

\begin{abstract}
Phonon anharmonicity is ubiquitous in real materials and is crucial for understanding thermal properties and phase stability. In this work, we show that anharmonic phonon modes can be obtained by maximizing their vibration stability during fitting the atomic trajectory. We prove that all information about these quasiparticles is contained in two small correlation matrices $\mathcal{S}$ and $\mathcal{Q}$, which can be constructed directly from molecular dynamics simulations. Based on these matrices, we proposed an optimization scheme, which allows us to efficiently determine temperature-dependent phonon modes along with their frequencies and lifetimes. We verified this method by applying it to silicon and cubic CaSiO$_3$, where it successfully captured their temperature-dependent phonon behaviors and the well-known phonon softening in cubic CaSiO$_3$. This theory provides a convenient tool for investigating phonon quasiparticles and can be extended to study other quasiparticles, such as electrons, holes, and magnons.
\end{abstract}

\maketitle

Phonons are quantized vibrational modes in crystalline solids, describing how atoms collectively oscillate around their equilibrium positions. These fundamental quasiparticles are essential in physics textbooks, playing a crucial role in our understanding of heat transfer, superconductivity, and thermoelectric effects. Modern theoretical studies determine phonon modes and frequencies through harmonic approximation, using density functional theory (DFT) with either perturbation theory (DFPT)~\cite{Zein1984,Baroni1987,Gonze1995,Baroni2001} or the finite displacement method~\cite{PHONON1997,PHON2009,YPHON2014,Phonopy2015}. The resulting phonons all have infinite life time. However, in real materials, phonons scatter with each other~\cite{Rollins1964PhononPhonon,delaire2011PhononPhonon,Cao2023PhononPhonon} and with other quasiparticles~\cite{fu2023PhononPhoton,Marini2024PhononElectron,marini2023PhononElectron,Sinha1962PhononMagnon}, resulting in finite lifetimes and modified material properties.

Over the past decades, researchers have developed various methods to study phonons beyond the harmonic approximation.   A natural method is to include phonon-phonon collisions using the third-order or even fourth-order force constants~\cite{Deinzer,Phono3py2015,Eriksson2019}.
This approach has effectively determined phonon lifetimes and thermal conductivity across various materials, including semiconductors~\cite{Broido2007,Lindsay2012,Xie2020}, rocksalt materials~\cite{Xia2020}, zincblende compounds~\cite{Debernardi1998,Xia2020,Ravichandran2020,Togo2023} and wurtzite compounds~\cite{Togo2023}, and two-dimensional crystals ~\cite{Feng2018,Tristant2019,Sun2023,Danayat2024}.
Although these perturbative approaches have proven successful, numerous non-perturbative methods have also emerged, such as the self-consistent harmonic approximation (SCHA)~\cite{Hooton01041955,koehler1966theory,errea2012isotope}, self-consistent {\it ab initio} lattice dynamics (SCAILD)~\cite{SCAILD2008}, stochastic self-consistent harmonic approximation (SSCHA)~\cite{PhysRevLett.111.177002,SSCHA2014,monacelli2021stochastic}, self-consistent phonon calculations (SCPH)~\cite{SCPH2015}, quantum self-consistent {\it ab initio} lattice dynamics (QSCHILD)~\cite{QSCAILD2021}, and use compressive sensing lattice dynamics~\cite{Zhou2019}.
Moreover, classical and {\it ab initio} molecular dynamics (MD) offers another powerful non-perturbative approach for studying phonon anharmonicity, as atomic trajectories naturally incorporate anharmonic effects to all orders. {  These classical trajectories can be utilized to extract renormalized vibrational modes, thereby providing crucial information about their quantum counterparts, the phonons~\cite{Sun2010QPA}.} The fluctuation-dissipation theorem can be used to obtain dynamical matrices from MD simulations~\cite{KONG20112201}. The temperature-dependent effective potential (TDEP) method~\cite{TDEP2013,zhou2018electron} and temperature dependent force constants~\cite{Esfarjani1,Esfarjani3,ALAMODE2014} can be extracted from AIMD simulations, enabling the calculation of anharmonic phonon properties. Similarly, the phonon quasiparticle approach (PQA) projects AIMD trajectories onto normal modes to directly obtain anharmonic phonon frequencies and lifetimes~\cite{PQA2009,PQA2014a,PQA2014b}. With all these methods yielding phonon quasiparticles via distinct approaches, a fundamental question arises: How accurate are these phonon quasiparticles in modeling the vibration of a real system?

\textbf{The key is vibration stability. In fact, phonon quasiparticles with inaccurate phonon modes exhibit less vibration stability when decomposing atom trajectories. Therefore, modes with maximized stability provide the most accurate representation of phonon quasiparticles. This principle inspires a concise and efficient method, developed in this study, to extract phonon quasiparticles from MD simulations. }

For a pure harmonic system with $n$ atoms in primitive cell, the $j$-th phonon mode at $\vec{k}$ point of Brillouin zone is $\varphi_{i\sigma}^{\vec{k},j}$ ($i=1, 2,..,n;\, \sigma=x,y,z$). There are 3$n$ phonon modes in total, forming a complete orthonormal basis set. Then, the mass-weighted atom trajectories or velocities in a MD simulation can be expanded as $\mathcal{R}_{i\sigma}^I(t)=Re\left(\sum_{\vec{k},j}c^{\vec{k},j}\varphi_{i\sigma}^{\vec{k},j} e^{\mathbf{i}(\vec{k}\cdot\vec{R}_I-\omega_{\vec{k},j}t )}\right)$, where $\omega$ is eigen frequency, $t$ is time, and $\vec{R}_I$ is the {  mass-weighted} displacement vector of the $I$-th cell.
The coefficient $c_{\vec{k},j}$ represents the weight of mode $\varphi^{\vec{k},j}$. Since $\mathcal{R}_{i\sigma}^I(t)$ contains phonon modes at all $\vec{k}$ points, we can filter out the modes at {  each} $\vec{k}$ by real space Fourier transformation:
\begin{equation}
	\begin{aligned}
		&f^{\vec{k}}_{i\sigma}(t)=\frac{1}{N}\sum_{I}\mathcal{R}_{i\sigma}^I(t)e^{-\mathbf{i}\vec{k}\cdot\vec{R}_I}\\
		&=\frac{1}{2}\sum_{j}c^{\vec{k},j}\varphi_{i\sigma}^{\vec{k},j} e^{-\mathbf{i}\omega_{\vec{k},j}t }+\frac{1}{2}\sum_{j}(c^{-\vec{k},j})^*(\varphi_{i\sigma}^{-\vec{k},j})^* e^{\mathbf{i}\omega_{-\vec{k},j}t }
   \end{aligned}\label{f}
\end{equation}
It turns out that the result contains both positive frequencies from $\vec{k}$ point and negative frequencies from $-\vec{k}$ point. To isolate contributions solely from the $\vec{k}$ point, we perform time Fourier transformation, remove negative frequencies, and apply inverse Fourier transformation, yielding:
\begin{equation}
	\begin{aligned}
		g^{\vec{k}}(t)&={  \frac{1}{2\pi}\int_0^{\infty}\left[\int_{-t_i}^{t_f} f^{\vec{k}}(\tau)e^{\mathbf{i}\omega\tau}d\tau\right] e^{-\mathbf{i}\omega t}d\omega } \\
    &=\frac{1}{2}\sum_{j}c^{\vec{k},j}\varphi^{\vec{k},j} e^{-\mathbf{i}\omega_{\vec{k},j}t }
	\end{aligned}\label{g}
\end{equation}
We have omitted the subscripts $i$ and $\sigma$ for simplicity, which makes $\varphi^{\vec{k},j}$ and $g^{\vec{k}}(t)$ column vectors with 3$n$ elements. {  In practical applications, smoothing is applied to the start and end points of $f^{\vec{k}}(t)$ to eliminate sharp cutoffs.}

Eq. \ref{g} also applies to real systems.
The key differences are that both modes and frequencies depend on MD temperature, and the coefficients $c^{\vec{k},j}$ become time-dependent due to mode scattering caused by anharmonicity.  The value $|\frac{dc^{\vec{k},j}(t)}{dt}|$ quantifies the vibration stability of mode $\varphi^{\vec{k},j}$. If inaccurate modes and the associated frequencies are used to expand MD trajectories, the coefficients $c^{\vec{k},j}(t)$ will vary more rapidly over time. This is clearer in a pure harmonic system, when MD trajectories are decomposed by non-eigenvectors, the normally invariant coefficients become time-dependent. This suggests that optimal phonon quasiparticles can be defined as those modes that minimize $|\frac{dc^{\vec{k},j}(t)}{dt}|$, equivalently speaking maximize stability.

Next, we describe the method for obtaining optimal phonon quasiparticles. After a MD simulation, the column vector $g^{\vec{k}}(t)$ is constructed directly from either mass-weighted atom trajectories or mass-weighted velocities {  using the first lines of Eq. \ref{f} and \ref{g}}. The time-dependent coefficient is then given by $c^{\vec{k},j}(t)=2(\varphi^{\vec{k},j})^+g^{\vec{k}}(t)e^{\mathbf{i}\omega_{\vec{k},j}t}$. For mathematical convenience, we minimize $|\frac{dc^{j}(t)}{dt}|^2$ rather than $|\frac{dc^{\vec{k},j}(t)}{dt}|$ to optimize phonon quasiparticles. Since the optimization was performed individually for each $\vec{k}$ point, for brevity, we omit the $\vec{k}$ superscript in the following discussions, unless needed for clarity.

\begin{figure}[t]
\includegraphics[width=0.5\textwidth]{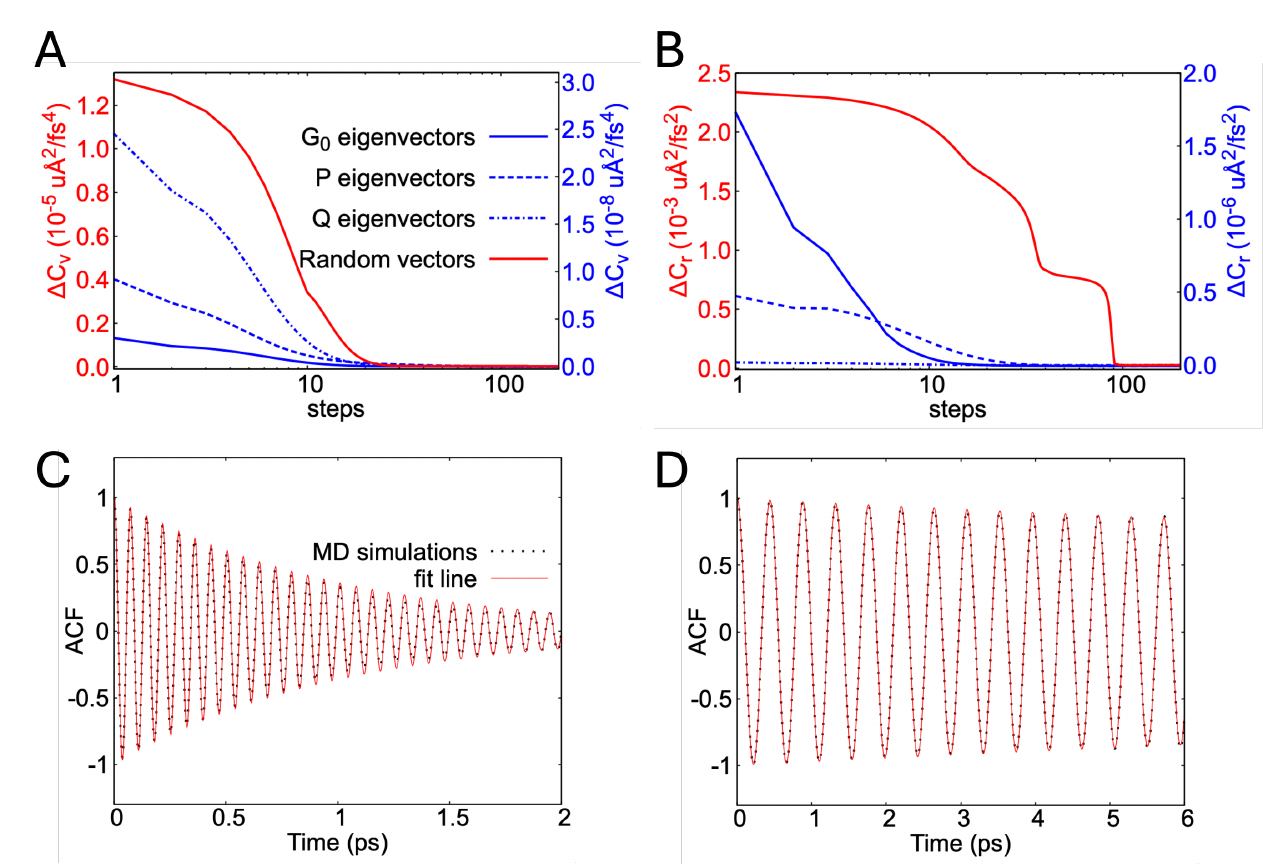}
\caption{Optimization process and ACF for silicon phonons. (A) $\Delta C_v$ obtained using mass-weighted velocities for different initial modes. (B) $\Delta C_r$ obtained using mass-weighted trajectories. In both panels, $\Delta C$ represents the difference between $C$ and its optimal value. (C) Normalized ACF of an optical phonon. (D) Normalized ACF of an acoustic phonon. All calculations performed at $\vec{k}=(1/5,1/5,1/5)$ at 500K. }
\label{Fig1}
\end{figure}

Starting from an initial guess of the mode set $\varphi^{j}$ ($j$=1,2,...,3$n$), we compute the sum of $|\frac{dc^{j}(t)}{dt}|^2$ and then take its time average:
\begin{equation*}
	\begin{aligned}
C=\frac{1}{t_f-t_i}\sum_j\int_{t_i}^{t_f} |\frac{dc^{j}(t)}{dt}|^2 dt=\sum_j(\varphi^{j})^+\mathcal{G}^{j}\varphi^{j}
	\end{aligned},
\end{equation*}
where $t_i$ and $t_f$ are the initial and final simulation times, and $\mathcal{G}^{j}=\mathcal{G}_0+\omega_{j}\mathcal{P}+\omega_{j}^2\mathcal{Q}$.  {  It quantitatively characterizes the overall instability of the guessed modes.} The matrices $\mathcal{G}_0$, $\mathcal{P}$, and $\mathcal{Q}$ here are independent of the initial guessed phonon modes. They are Hermitian and can be constructed solely from the MD trajectory as equal-time correlation matrices,

\begin{equation*}
	\begin{aligned}
\mathcal{G}_0&=\frac{4}{t_f-t_i}\int_{t_i}^{t_f} \frac{dg(t)}{dt}(\frac{dg(t)}{dt})^+ dt \,,\\
\mathcal{Q}&=\frac{4}{t_f-t_i}\int_{t_i}^{t_f} g(t)(g(t))^+ dt \,,\\
\mathcal{P}&=\mathbf{i}[(\mathcal{S})^+-\mathcal{S}] \,,\\
\mathcal{S}&=\frac{4}{t_f-t_i}\int_{t_i}^{t_f} \frac{dg(t)}{dt}(g(t))^+ dt \,.
	\end{aligned}
\end{equation*}
{ $\mathcal{G}_0$, $\mathcal{Q}$, and $\mathcal{S}$ denote the velocity-velocity, displacement-displacement, and velocity-displacement correlation matrices, respectively, while $\mathcal{P}$ represents the imaginary part of $\mathcal{S}$. If $g(t)$ is constructed using velocities, these matrices effectively characterize the correlations between accelerations and velocities. }
Once these matrices are constructed, the minimization of $C$ is straightforward. The optimal frequencies satisfy the relation  $\frac{\partial C}{\partial\omega_{j}}=0$. Then we get
\begin{equation}\label{omega}
   \omega_{j}=-\frac{P_{jj}}{2Q_{jj}}=-\frac{Im(S_{jj})}{Q_{jj}},
\end{equation}
where $Im(S_{jj})$ pickes the imaginary part of $S_{jj}$, the matrix elements $P_{ij}=(\varphi^{i})^+  \mathcal{P} \varphi^{j}$, $Q_{ij}=(\varphi^{i})^+  \mathcal{Q} \varphi^{j}$, and  $S_{ij}=(\varphi^{i})^+  \mathcal{S} \varphi^{j}$.

The optimization of phonon modes requires a more complex process involving a series of rotations in Hilbert space, { $\Psi \rightarrow \Psi U$ with matrix $\Psi=\left[\varphi^{1},\varphi^{2}, ..., \varphi^{3n}\right]$}, where the matrix $U$ is unitary. This rotation results in a variation of $C$:
\begin{equation*}
	\begin{aligned}
\delta C =\delta Tr\left( PU\Omega U^+\right)+\delta Tr\left(QU\Omega^2 U^+\right)
	\end{aligned}.
\end{equation*}
The $Tr$ notation here represents the trace of a matrix. The matrix $\Omega$ is diagonal, with its diagonal elements corresponding to the mode frequencies. As we have shown, these frequencies depend on the phonon modes, and are therefore dependent on $U$. This variation equation leads us to:
\begin{equation*}
	\begin{aligned}
\frac{\delta C}{(\delta U)_{ij}}=(\Omega_{jj}-\Omega_{ii} )P_{ji}+(\Omega_{jj}^2-\Omega_{ii}^2 )Q_{ji}
	\end{aligned}.
\end{equation*}
The derivation of these two equations is somewhat tedious, we present them in the supplementary materials for completeness.

Let's define a matrix $B$, whose elements are
\begin{equation}
	\begin{aligned}
	B_{ij}=\left(\frac{\delta C}{\delta U_{ij}}\right)^*&=(\Omega_{jj}-\Omega_{ii} )P_{ij}+(\Omega_{jj}^2-\Omega_{ii}^2 )Q_{ij} ,\\
	{   B}&{ =[P,\Omega]+[Q,\Omega^2]} .
	\end{aligned}\label{opt}
\end{equation}
This represents the fastest direction for increasing $C$. { Note that the matrices $P$ and $Q$ are not the raw correlation matrices defined earlier, but their mode-projected counterparts, defined as $P=\Psi^\dagger \mathcal{P} \Psi $ and $Q=\Psi^\dagger \mathcal{Q} \Psi $.}
The complex conjugate operation ensures that $\delta C$ remains real. To minimize $C$, we define the unitary transformation as $U=e^{-\gamma B}$, where $\gamma$ is the step length of rotation in Hilbert space. This value should be small enough to maintain stability in the minimization process. In practical applications, we can use the approximation $U=1-\gamma B$. By repeatedly performing these calculations until $B$ is zero, we ultimately obtain the optimal phonon modes and their frequencies.

\begin{figure*}[t]
	\includegraphics[width=1.0\textwidth]{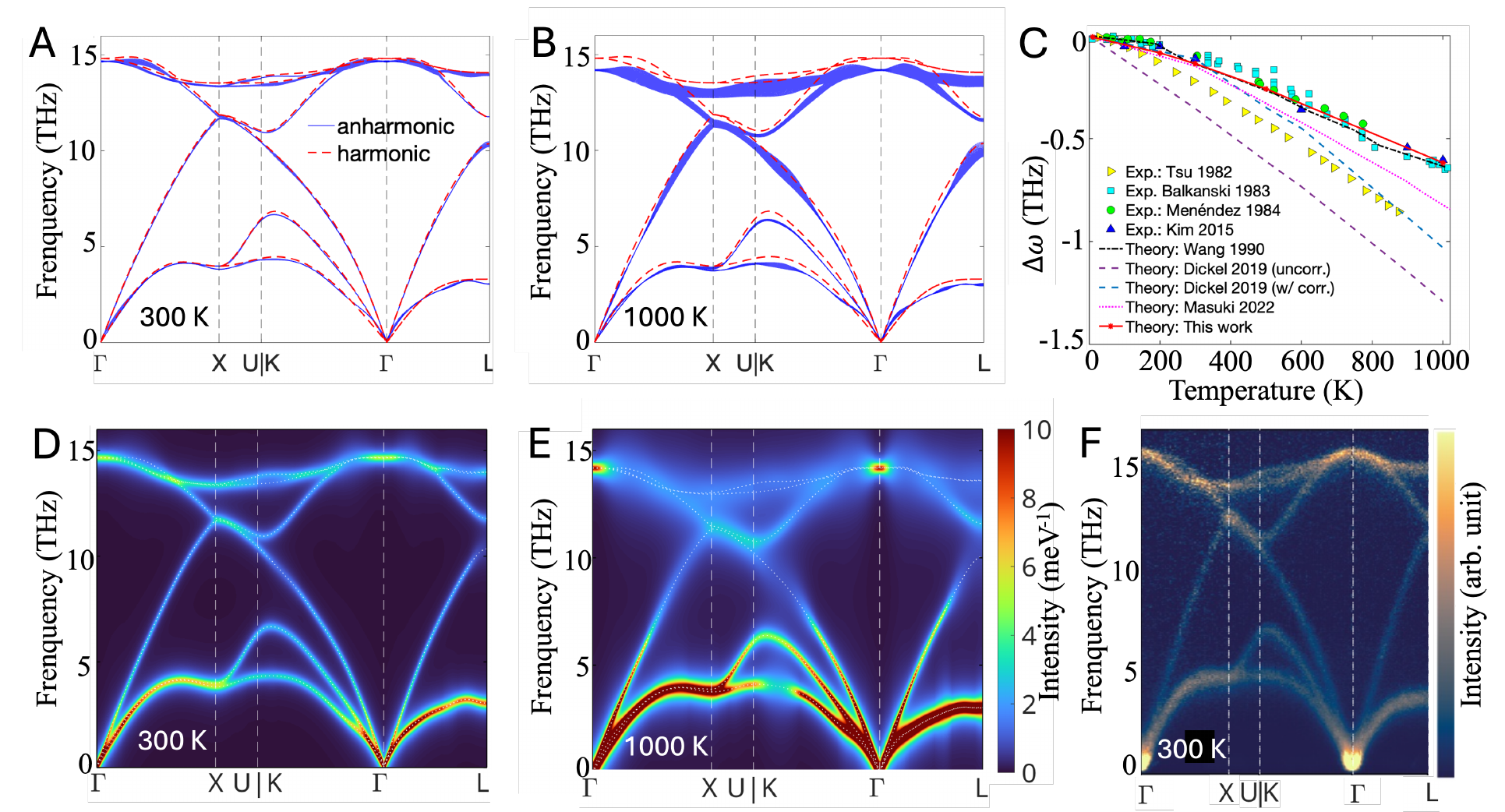}
	\caption{{  Temperature-dependent phonon properties of silicon. (A, B) Harmonic (red lines) and anharmonic (blue lines) phonon dispersions at (A) 300 K} and (B) 1000 K. The linewidths are represented by the thickness of the curves. {   (D, E) The corresponding phonon scattering functions. (C) Thermal shift of the LO phonon at the $\Gamma$ point, compared with experimental results from Ref. \cite{Raphael1982SiExp,Balkanski1983SiExp,Menendez1984SiExp,silicon2015} and theoretical results from Ref.\cite{silicon1990,Dickel2019SiTheory,Masuki2022SiTheory} . (F) Experimental phonon scattering function at 300 K. Panel (F) is reprinted from Kim et al., PNAS 115, 1992 (2018), published under the PNAS license.}}
	\label{Fig2}
\end{figure*}

Having established this theoretical framework, we apply the optimization method to a real material, silicon. To enhance computational efficiency, we performed a 5$\times$5$\times$5 supercell MD simulation at 500 K and 1000 K over 50 ps using the first-principles machine learning force field in VASP code~\cite{vasp}. The simulation begins with a 4 ps canonical (NVT) ensemble MD to achieve thermal equilibrium, followed by a transition to the microcanonical (NVE) ensemble MD.
The NVE ensemble is suitable here because it eliminates artificial interactions between phonon modes and the heat bath. Subsequently, function $g(t)$ was constructed based on the mass-weighted trajectories and velocities.  Six phonon modes were randomly generated and orthogonalized using the Gram-Schmidt procedure. As shown in { Fig. 1~A}, the initial value of $\Delta C$ for $\vec{k}=(1/5,1/5,1/5)$ at 500 K based on mass-weighted trajectories is substantially large but decreased rapidly during optimization. The value of $\Delta C$ based on mass-weighted velocities has the same trend. These results demonstrate the high efficiency of the optimization method.

The efficiency of the optimization process can be further enhanced by selecting better initial modes.
The harmonic phonon modes are a viable option. Additionally, the peak values from the Fourier transformation of Eq. \ref{f} serve as effective initial modes. In fact, the eigenvectors of $\mathcal{G}_0$, $\mathcal{P}$, and $\mathcal{Q}$ provide even more convenient alternatives. Take $\mathcal{Q}$ as an example, combining its definition and the expression of $g(t)$ yields,
\begin{equation*}
	\begin{aligned}
		\mathcal{Q}&=\frac{1}{t_f-t_i}\sum_{i,j}\varphi^{i} (\varphi^{j})^+\int_{t_i}^{t_f} c^i(t)(c^j(t))^* e^{\mathbf{i}(\omega_j-\omega_i)t}dt\\
         &\approx\frac{1}{t_f-t_i}\sum_{i,j}\varphi^{i} (\varphi^{j})^+\delta_{\omega_{i},\omega_{j}}\int_{t_i}^{t_f}  c^{i}(t)(c^{j}(t))^*dt.
	\end{aligned}
\end{equation*}
The integration in the first line consists of a slowly varying function $c^i(t)(c^j(t))^*$ and an exponential function $e^{\mathbf{i}(\omega_j-\omega_i)t}$. The latter oscillates quickly if $\omega_j\ne \omega_i$, making the integration negligible and producing the delta function in the last line.  This form indicates that the eigenvectors of $\mathcal{Q}$ are isolated anharmonic phonon modes and the linear combinations of degenerated anharmonic phonon modes, while the eigenvalues are their weights. The same argument applies to $\mathcal{G}_0$ and $\mathcal{P}$ as well, though with different eigenvalues.  We should caution here that due to the finite MD simulation time and the phonon anharmonicity, the last line is only an approximation. As a result, the eigenvectors do not precisely represent the anharmonic phonon modes, which requires optimization. The optimization results using the eigenvectors of $\mathcal{G}_0$, $\mathcal{P}$, and $\mathcal{Q}$ as initial modes are displayed as red lines in { Fig. 1~A and B}. We can see that the initial values of $\Delta C$ are three orders of magnitude smaller than those obtained from random initial modes. Therefore, these initial modes are very effective, allowing us to get more accurate anharmonic phonons within the same number of optimization steps.

We now proceed to calculate the phonon linewidth, which is inversely proportional to the phonon lifetime~\cite{stefanucci2013}. Previous studies have shown that phonon linewidth can be obtained from the mode-projected velocity autocorrelation function (ACF)~\cite{PQA2014a,PQA2014b}. Similarly, in this work, phonon linewidth can also be derived from the ACF of mode-projected $g(t)$ function, $g_j(t)=(\varphi^j)^+g(t)$:
\begin{equation}
	\begin{aligned}
		<g_j(0)\cdot g_{j}(t)>&=\frac{1}{t_f-t_i}\int_{t_i}^{t_f} g_j^*(\tau)g_j(t+\tau)d\tau \\
		&=Q_{jj}e^{-i\omega_{j}t}e^{-\Gamma_{j}t}.
	\end{aligned}\label{gamma_exp}
\end{equation}
{  The term $e^{-i\omega_{j}t}$ represents the mode’s coherent oscillation at its frequency $\omega_j$, while $e^{-\Gamma_{j}t}$ captures dissipative processes, with $\Gamma_j$ quantifying the rate of energy loss. Together, they describe the reversible dynamics and irreversible energy loss, enabling the extraction of key dynamical parameters.} The frequency and lifetime can be obtained from the time derivative of the ACF:
\begin{equation}
	\begin{aligned}
		-i\omega_j-\Gamma_j=\frac{<g_j(0)\cdot \dot{g}_{j}(0^+)>}{Q_{jj}}=\frac{S_{jj}}{ Q_{jj} }.
	\end{aligned} \label{gamma}
\end{equation}
Eq. \ref{omega} represents the imaginary part of this equation. For the real part, the value of $S_{jj}$ must be constructed using $\dot{g}_j$ at a time slightly later than $g_j$. This is because the exponential function in Eq. \ref{gamma_exp} is only valid for $t>0$. This requirement increases the calculation complexity, making it simpler to obtain phonon lifetime by fitting the ACF using Eq. \ref{gamma_exp}.

\textbf{The phonon quasiparticle calculation procedure begins with constructing $\mathcal{S}$ and $\mathcal{Q}$ from MD trajectories. Next, $\mathcal{Q}$ is diagonalized to obtain its eigenvectors, which serve as the initial phonon modes. These modes are then optimized by applying Eqs. \ref{omega} and \ref{opt}, yielding the optimal phonon modes and frequencies. Finally, the linewidths are calculated from ACF.}

The autocorrelation function for silicon are shown in { Fig. \ref{Fig1}~C and D. Panel C} displays an optical phonon with a frequency of 13.9 THz and lifetime of 0.8 ps. { Panel D} shows an acoustic phonon with a frequency of 2.3 THz and a significantly longer lifetime of 36.1 ps.

For each $\vec{k}$ point, these equations enable us to determine the optimal phonon modes, frequencies, and their linewidths. Let's proceed to calculate the phonon dispersions. The dynamical matrix can be constructed as $D_{\vec{k}}=\varPhi_{\vec{k}}\widetilde{\Omega}_{\vec{k}}^2\Phi_{\vec{k}}^+$, where $\varPhi_{\vec{k}}=[\varphi^{\vec{k},1},\varphi^{\vec{k},2},...,\varphi^{\vec{k},3n}]$, and $\widetilde{\Omega}_{\vec{k}}$ is a complex matrix with elements $\widetilde{\omega}^j_{\vec{k}}=\omega^j_{\vec{k}}-\mathbf{i}\Gamma^j_{\vec{k}}$. The force constant matrix is given by $F_{\vec{R}}=\frac{1}{N}\sum_{\vec{k}}D_{\vec{k}}e^{\bold{i}\vec{k}\cdot\vec{R}}$, with the sum performed over all $\vec{k}$ points in the first Brillouin zone compatible with the supercell. Then the dynamical matrix at $\vec{k}'$ points along the high-symmetry line of the first-Brillouin zone is $D_{\vec{k}'}=\sum_{\vec{R}}F_{\vec{R}}e^{-\bold{i}\vec{k}'\cdot\vec{R}}$. Finally, diagonalizing these matrices produces the phonon dispersions, linewidths, and modes. 

{  In this scheme, the imaginary part of the eigenfrequencies is interpreted as originating from finite lifetimes. However, in an unstable system, soft phonon modes may also yield purely imaginary eigenfrequencies. To distinguish between these two sources, we employ the following effective method: if the imaginary part of the frequency exceeds the real part, we redefine the frequency as purely imaginary. In such cases, the mode cannot complete even half a period of oscillation within its lifetime; consequently, the lifetime becomes ill-defined, implying that the imaginary part arises from the instability of the phonon. We represent these modes as negative frequencies in the phonon dispersion plots. This approach is a natural generalization of the imaginary frequencies in harmonic phonon dispersions.}

\begin{figure*}[t]
	\includegraphics[width=1\textwidth]{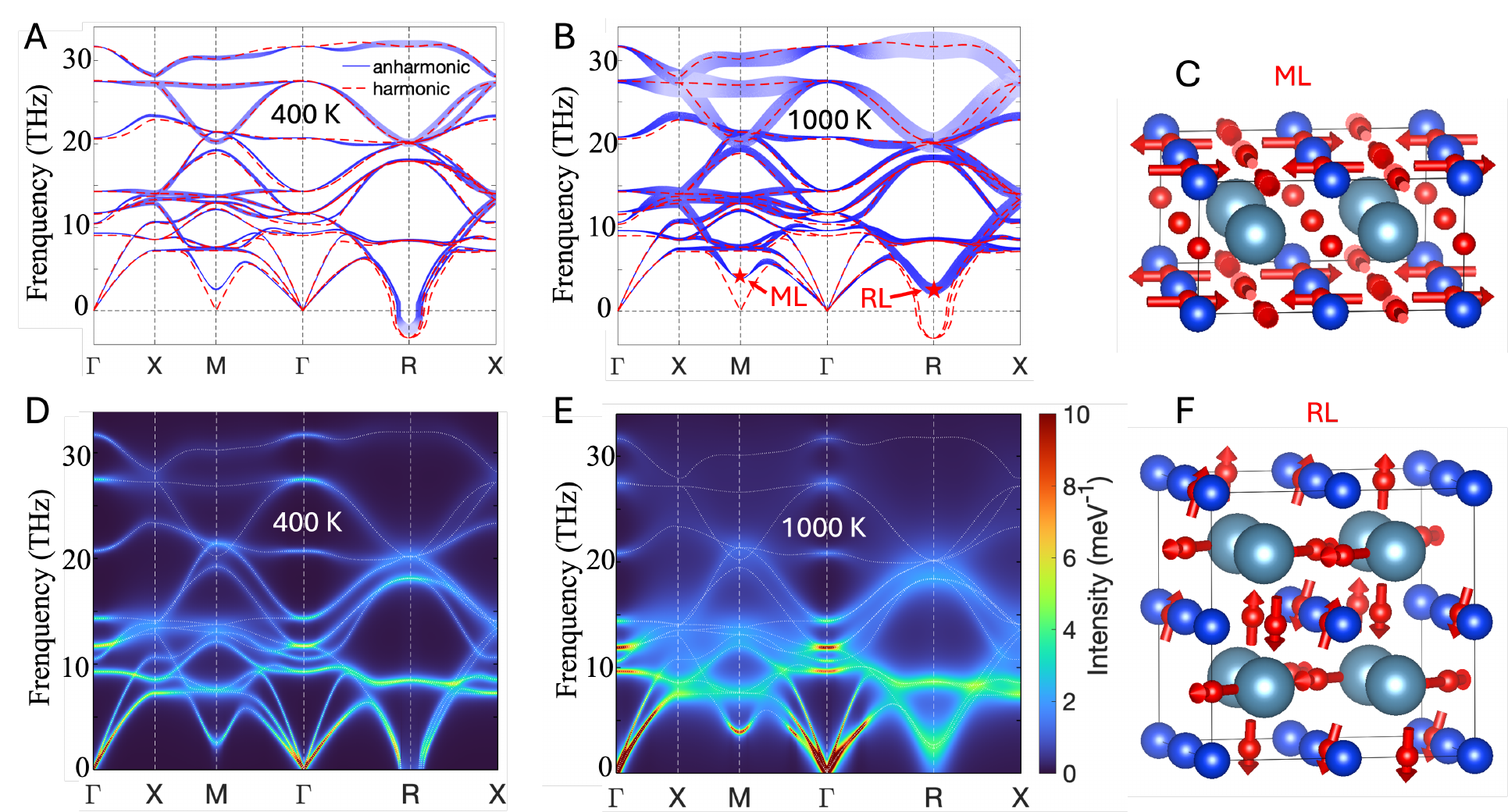}
	\caption{{  Harmonic (red lines) and anharmonic (blue lines) phonon dispersions} for CaSiO$_3$ at (A) 400 K and (B) 1000 K. The line thickness represents the phonon linewidths. Panels (D, E) display the corresponding {  phonon scattering functions}, while panels (C, F) show the soft phonon modes ML and RL at 1000 K. }
	\label{Fig3}
\end{figure*}

{  The calculated phonon properties for silicon are shown in Fig. \ref{Fig2}}. MD simulations were performed { at 300 K and 1000 K}, with the resulting phonon dispersions displayed in { Fig. \ref{Fig2}~A and Fig. \ref{Fig2}~B}, respectively. These dispersions closely match the harmonic phonon dispersions, indicating that anharmonic effects are weak in this system. The three optical phonons display larger linewidths than the acoustic phonons. Due to larger vibration amplitudes causing stronger anharmonic effects, the phonon linewidths at 1000 K are significantly larger than those at {  300 K}. Furthermore, compared to the phonon dispersions at {  300 K}, all phonon frequencies at 1000 K are reduced, agree with previous theoretical and experimental studies. {  The thermal shift of the longitudinal optical (LO) phonon frequency at $\Gamma$ point as a function of temperature is plotted at Fig. \ref{Fig2}~C, along which is the experimental results~\cite{Raphael1982SiExp,Balkanski1983SiExp,Menendez1984SiExp,silicon2015} and previous theoretical results~\cite{silicon1990,Masuki2022SiTheory,Dickel2019SiTheory}. The experimental data from Refs.~\cite{Balkanski1983SiExp,Menendez1984SiExp,silicon2015} exhibit remarkable consistency, confirming their reliability. Our theoretical predictions, as well as the results derived from the velocity-velocity correlation of raw MD trajectories by Wang et al.~\cite{silicon1990}, are in excellent agreement with these experimental benchmarks.}

With the phonon frequencies and linewidths calculated along the high-symmetry lines, we can now plot the {  phonon scattering functions}, which can be directly compared with experimental results. Based on the Bose-Einstein distribution and the standard Lorentzian function of spectral lines, the { Stokes dynamical structure factor } at $\vec{k}$ point can be written as
$$
W(\omega)=\frac{1}{\pi}\sum_{j}{\left(\frac{1}{e^{\hbar \omega_j/kT}-1}+1\right)}\frac{\Gamma_j}{(\omega-\omega_j)^2+{\Gamma_j^2}},
$$
where $\hbar$ is the reduced Planck constant, $k$ is the Boltzmann constant, and $T$ is temperature. The calculated results are shown in {  Fig. \ref{Fig2}~D and E}. At {  300 K, the phonon scattering functions are sharp and well-resolved, indicative of weak phonon interactions. Conversely, at 1000 K, the phonons have more chance  to collision with each other, then the functions display significant dispersive broadening. The calculated results at 300 K show excellent agreement with inelastic neutron scattering measurements~\cite{Kim2018} in Fig. \ref{Fig2}~F. Both experimental and theoretical results reveal higher intensity for acoustic phonons below 5 THz and optical phonons around 15 THz, while exhibiting weaker intensity in the intermediate frequency region.}

Let's continue to study a more complex material, the perovskite CaSiO$_3$ at high pressure.  It has strong anharmonic effect of phonons~\cite{PQA2014b,Renata2021,CaPv2024}, and is a main component of pyrolitic lower mantle~\cite{mcdonough1995composition,valencia2017influence}. The harmonic phonon dispersions of cubic CaSiO$_3$ show imaginary frequencies at all pressures, therefore, it is unstable and can change to more stable phases such as tetragonal or orthorhombic structures.  Experiments have shown that the cubic  CaSiO$_3$  can be stabilized by high teperature and pressure~\cite{KOMABAYASHI2007564,kurashina2004phase,ono2004phase}. Many researchers have studied its anharmonic phonons, and the study of this material is still active~\cite{thomson2019seismic,CaPv2021,CaPv2022} in geoscience.

Unlike silicon, cubic CaSiO$_3$ is a typical polar material. In such systems, the transverse optical (TO) phonon modes and LO phonon modes split {  at $\Gamma$ point} due to long-range Coulomb interactions. Therefore, after obtaining the effective force constants, we need to apply corrections~\cite{Pick1970} based on Born effective charges and the dielectric tensor. The lattice parameter of the unit cell was set to 3.45 Å, corresponding to the pressure at the top of the lower mantle. The calculated phonon dispersions of cubic CaSiO$_3$ at 400 K and 1000 K are shown in { Fig. \ref{Fig3}~A and B}. The resulting phonon spectral functions are shown in {  Fig. \ref{Fig3}~D and E}. {  Both the lowest phonon modes at the $M$ and $R$ points soften at low temperatures. The three lowest phonon branches around the $R$ point exhibit imaginary frequencies in Fig. \ref{Fig3}~A, demonstrating that cubic CaSiO$_3$ is unstable at 400 K. Conversely, their frequencies are positive in Fig. \ref{Fig3}~B, indicating that the system is stable at 1000 K. This result is consistent with previous X-ray diffraction (XRD) measurements~\cite{KOMABAYASHI2007564,ono2004phase,kurashina2004phase}.} The lowest modes at the $M$ and $R$ points, shown in {  Fig. \ref{Fig3}~C and F}, are characterized by the vibration of O atoms.

\textbf{The use of compact matrices ensures the efficiency of this method, while the absence of approximations guarantees its accuracy. This paradigm is broadly applicable to other systems governed by their own quasiparticles. For example, in the case of magnons, compact matrices can be derived from spin trajectories calculated using the Landau-Lifshitz-Gilbert equation. Similarly, for electrons and holes, compact matrices or kernel functions can be extracted from time-dependent DFT results. The corresponding quasiparticles can then be systematically determined from these matrices or kernel functions.}

In summary, this work defines optimal phonon quasiparticles as those that maximize their stability. We demonstrate that all information about these quasiparticles is stored in two simple matrices, $\mathcal{S}$ and $\mathcal{Q}$, which can be constructed directly from MD trajectories or velocities. The eigen vectors of $\mathcal{Q}$ can be used as the initial phonon modes. Then the optimal phonons and their corresponding frequencies and lifetimes can be obtained from an optimization scheme. Tests on silicon show weak anharmonicity with temperature-dependent shifts, while cubic CaSiO$_3$ exhibits strong anharmonic effects, being unstable at 400 K but stabilized at 1000 K. These tests show the reliability of this method. This theory and optimization method provide a clear picture of phonon quasiparticles in real materials, offering a convenient tool for their investigation. Its core principles are readily transferable, paving the way for the study of a broad spectrum of other quasiparticles.

\begin{acknowledgments}
	W. Li and Y. Lu contributed equally to this work. We thank T. Sun, D. Zhang, F. Zhou, Z.-K. Liu, A. Togo, and S. S. Naghavi for useful discussions. This work was supported by the National Natural Science Foundation of China under Grants 12374054 and 12022415. We acknowledge the computing resources of the Tencent TEFS platform (https://tefscloud.com).
\end{acknowledgments}


%

\end{document}